Dramatic relativistic and magnetic Breit effects for the superheavy reaction Og +3Ts$_2$ → OgTs6: Prediction of atomization energy and the existence of the superheavy octahedral Oganesson hexatennesside OgTs$_6$

Gulzari L Malli [a]

Department of Chemistry, Simon Fraser University

8888 University Dr., Burnaby, B.C; Canada V5A 1S6.

Martin Siegert, [b]

Research Computing, IT Services, Simon Fraser University

8888 University Dr., Burnaby, BC; Canada V5A 1S6

Luiz Guilherme M. de Macedo [c]

Universidade Federal de São João del-Rei, UFSJ

Campus Centro-Oeste Dona Lindu - CCO, UFSJ

Rua Sebastião Gonçalves Coelho, 400

Bairro Chanadour - Divinópolis, MG - Brazil 35.501-296

Walter Loveland [d]

Department of Chemistry, Oregon State University

Corvallis, Oregon, OR 97331-4003, USA

[a] Corresponding author email: malli@sfu.ca

[b] email:siegert@sfu.ca

[c] email: lgm@ufsj.edu.br

[d] email:lovelanw@onid.orst.edu

**Abstract** Our gargantuan *ab initio* all-electron fully relativistic Dirac–Fock (DF), nonrelativistic (NR) Hartree-Fock (HF) and Dirac-Fock-Breit-Gaunt(DFBG) molecular SCF calculations for the superheavy octahedral Oganesson hexatenniside $OgTs_6$ predict atomization energy ($A_e$) of 9.49, -5.54 and 9.37 eV, at the optimized Og-Ts bond distances of 3.35, 3.44 and 3.36 Å, respectively. There are *dramatic* effects of relativity for the atomization energy of $OgTs_6$ (with seven superheavy elements and 820 electrons) of ~ 15.0 eV each at both the DF and DFBG levels of theory, respectively. Our calculated energy of reaction for the titled superheavy reaction $Og + 3Ts_2 \rightarrow OgTs_6$ at the DF, NR and, DFBG levels of the theory is 6.33, 8.81, and 6.26 eV, respectively. Mulliken analysis as implemented in the DIRAC code for our DF and NR calculations (using the dyall.cv4z basis) yields the charges $Og^{+0.60}$ and $Og^{+0.96}$, respectively on the central Og atom indicating that our relativistic DF calculations predict octahedral $OgTs_6$ to be less *ionic*. However, due caution must be used to interpret the results of Mulliken's population analysis, which is highly basis set dependent.

# 1 Introduction

A fundamental research area in chemistry is the discovery of new superheavy elements (SHE) (with Z> 103) and the investigation of their chemical properties. Recently an unparalleled event took place in the history of SHE with the discovery of four superheavy elements (SHE) [1, 2] with Z=113-118 which has resulted in the completion of the 7th row of the periodic table. The two heaviest SHE Tennessine Ts (Z=117) and Oganesson Og (Z=118) have been placed in the periodic table under the halogen At (Z=85) and the noble gas Rn (Z=86), because Ts and Og are expected to resemble their lighter congeners At and Rn, respectively. It is well-known that serious challenges confront both the experimentalists and the theoreticians working in the research area of SHE. Experimentally, the synthesis of SHE depends upon nuclear fusion reactions with heavy-ion beams using very rare targets of isotopes ($^{248}$Cm, $^{249}$Bk, etc) which are available at present only at a very few research laboratories in the world. Moreover, half-lives of the product species decrease very rapidly so that on average each synthesized atom has decayed before a new one is made. This has been dubbed as "one-atom-a-time or "single atom chemistry" which poses serious problems.

However, during the last decade, the experimentalists have successfully carried out state-of-the-art experiments for the chemical investigation of numerous SHE in various laboratories around the world [2-10].

Theoretical investigation of SHE and their compounds has its challenges which arise mainly due to significant relativistic effects in systems of SHE because of the very large nuclear charge (Z>103). This can seriously affect the dynamics of electrons of the SHE as with the increase of Z, the velocity (v) of an electron of SHE (With Z> 103) may approach very close to the velocity of light (c). Therefore, Schrodinger's nonrelativistic treatment for the electronic structure of SHE may not be appropriate. Secondly, with the enormous number of electrons in the SHE and their systems, the ubiquitous electron correlation should be included in the calculations for systems of SHE. It is well-established that Dirac's relativistic treatment for many-electron systems should be more appropriate for the systems of SHE. The LCAS/MS (where a molecular spinor MS is written as Linear Combination of Atomic Spinors (LCAS)) Dirac-Fock (DF) SCF theory for closed-shell molecules was developed [11] by Malli and Oreg in 1975 and for open-shell molecules by Malli [12] in 1980 which we have used extensively [13-23] to investigate the effects of relativity in the chemistry of heavy and SHE using the MOLFDIR [24] and the DIRAC [25] codes (hereafter referred to as the code ). Our goal in this paper is to investigate the relativistic DF and magnetic Breit effects for the atomization energy of Og $(Ts)_6$ and energy of reaction for the superheavy reaction: Og +3$Ts_2$ →$OgTs_6$. This system involves seven heaviest SHE with 820 electrons, and so all-electron DF, DFBG, and even NR HF calculations for this system would be *gargantuan*.

## 2 Theoretical Remarks

We present here only a brief outline of the Dirac-Fock-Breit-Gaunt (DFBG) treatment for molecules [and] refer the reader for further details to the DIRAC code [25] which we have employed for all of our calculations at various levels of theory. The approximate relativistic Dirac-Fock-Coulomb Hamiltonian ($H_{DC}$) for an N-electron molecular system containing n nuclei, under the Born-Oppenheimer approximation (omitting the nuclear repulsion terms which are constant for a given molecular configuration) can be written (in atomic units) [11]:

$$H_{DC} = \sum_{i=1}^{N} H_D(i) + \sum_{i<j} \frac{1}{r_{ij}} \qquad (1)$$

In Eq. (1), the $H_D(i)$ consists of the Dirac's kinetic energy operator, mass-energy and nuclear attraction of the i-th electron and has the well-known expression, viz;

$$H_D(i) = c\alpha_i \cdot p_i + (\beta_i - 1)c^2 + V_{nuc} \qquad (2)$$

The rest mass energy of an electron has been subtracted in Eq.(2), and the potential $V_{nuc}$ due to n *finite nuclei* of the molecular system is taken to be the sum of their nuclear potentials viz.; $V_{nuc} = \sum_n V_n$, and for molecular systems involving heavy atoms (with Z > 70), a finite nuclear model is invariably used. We shall use the Gaussian nuclear model [17] in which a single Gaussian function is used for each nuclear charge distribution. The instantaneous Coulomb repulsion between the electrons is treated *nonrelativistically* in the Dirac-Coulomb Hamiltonian and the magnetic and retardation corrections to it are generally included by perturbation theory. The N-electron wavefunction Φ for the closed-shell molecular system is taken as a single Slater determinant (SD), also called an antisymmetrized product (AP) of one-electron 4-component molecular spinors (MS) [11] viz.:

$$\Phi = (N!)^{-1/2} | \phi_1(1) \phi_2(2) \phi_3(3) \ldots \phi_N(N) |. \quad (4)$$

The molecular spinors (MS) $\phi_i$ are generally taken to form an orthonormal set and can be constructed to transform like the *extra* or *additional irreducible representations* (EIR) of the double symmetry group of the molecule under investigation.

The energy expectation value E can then be written as:

$$E = \langle \Phi | H | \Phi \rangle / \langle \Phi | \Phi \rangle \quad (5)$$

The molecular spinors $\phi_i$ are expressed in terms of the large and small components

$$\phi_i^X = \sum_{q=1}^{n} C_{iq}^X \chi_q^X, \quad X = L \text{ or } S.. \quad (6)$$

The $\chi_q^L$ and $\phi_i$ can be symmetry adapted; however, we shall ignore the double group symmetry labels. for the molecular spinors. The basis spinors $\chi_q^X$ will be constrained to obey the kinetic balance relation [26] viz,

$$\chi_q^S = (\sigma \cdot p) \chi_q^L. \quad (7)$$

Then following Malli and Oreg [11], the Dirac-Hartree-Fock-Roothaan (DHFR) or relativistic Hartree-Fock-Roothaan (RHFR) SCF equations for closed-shell molecules can be written as:

$$Fc_i = \varepsilon_i sc_i, \quad (8)$$

Where in Eq. (8), F is the Dirac- Fock matrix operator, the $\varepsilon_i$ is the orbital energy of the molecular spinor (MS) $\phi_i$ and S is the overlap matrix.

The Breit interaction [27] consisting of the magnetic and retardation terms was proposed to remedy partially the above-mentioned defect of $H_{DC}$ and the addition of the Breit interaction ($B_{ij}$) to $H_{DC}$ leads to the Dirac-Coulomb-Breit (DCB) Hamiltonian $H_{DCB}$, which has the form:

$$H_{DCB} = H_{DC} + \sum_{i<j} B_{ij}. \quad (9)$$

The $B_{ij}$ in Eq. (9) is usually written as:

$$B_{ij} = -\frac{1}{2}\{(\alpha_i \cdot \alpha_j)r_{ij}^{-1} + (\alpha_i \cdot r_{ij})(\alpha_j \cdot r_{ij})\, r_{ij}^{-3}\} \qquad (10)$$

Twice the first term in Eq. (10) called the magnetic or Gaunt interaction is the dominant part of Breit interaction; the retardation term is ~ 10% of the Gaunt interaction and, in general, the contribution of Breit interaction is fairly marginal compared to the Coulomb interaction term. The use of $H_{DC}$ plus the Gaunt Hamiltonian as the starting point for variational molecular calculations leads to the Dirac-Fock-Breit-Gaunt (DFBG) SCF equations. The Dirac-Fock-Breit-Gaunt matrix operator occurring in the DFBG SCF equations involves the matrix elements of the magnetic and Coulomb interactions. The expressions for the matrix elements of the magnetic Breit or Gaunt interaction are given in Malli and Oreg [11].

## 3 Dirac–Fock-Breit-Gaunt, Dirac-Fock, and NR Hartree-Fock calculations for Og $(Ts)_6$ and $Ts_2$

There are neither ab initio all-electron 4-component relativistic DF nor the corresponding NR Hartree-Fock calculations available for the superheavy $OgTs_6$ with seven superheavy atoms and 820 electrons. Needless to say that DF, NR HF and Dirac-Fock-Breit-Gaunt (DFBG) calculations for $OgTs_6$ would be gargantuan and would require a state-of-the-art supercomputer facility. It should be remarked that the magnetic Gaunt interaction can be included at the SCF stage in the DIRAC code [25] and therefore all-electron Dirac-Fock (DF), Dirac-Fock-Breit-Gaunt (DFBG), and the corresponding nonrelativistic (NR) Hartree-Fock calculations for the octahedral $OgTs_6$ were performed with the Dirac [25] code using the dyall.cv4z basis for Og and Ts, which are available within the Dirac website [25]. The total number of primitive gaussians used in our calculations is 11508 with 3570 L (large component) and 7938 S (small components) and the large component basis set for Og and Ts is 35s 35p 24d 16f 3g 1h. The kinetic balance [26] constraints as well as, the gaussian nuclear model as implemented in the code [25] are employed in all of our calculations. Our calculated total DF and DFBG energies using the above-mentioned basis and atomic masses used in the DIRAC code for Og are –54807.9332 and -54698.3124 au, while for Ts our calculated DF and DFBG energies are -53494.0590 and -53388.2300 au, respectively. The LEVY-LEBLOND option was used for all the corresponding NR HF calculations and our calculated total NR HF energies for Og and Ts are -46320.4278 and -45396.6978 au, respectively. All of our DFBG, DF, and NR HF calculations for the atoms Og and Ts, the diatomic $Ts_2$, and the polyatomic $OgTs_6$ were carried out using the above-mentioned appropriate basis with the Dirac code[25]. However, it should be pointed out that all of our DFBG, DF, and NR calculations for the octahedral $OgTs_6$ are performed with the Dirac code using the $D_{2h}$* double group. Geometry optimization for the octahedral $OgTs_6$ and the diatomic $Ts_2$ were carried out automatically using the Dirac code [25]. Mulliken population analysis [28] as implemented in the Dirac code [25] was carried out to obtain the charges on the Og and Ts atoms; however, it should be stressed, that this analysis is highly basis set dependent [29] and its results should be used with due caution. The results of our NR, DFBG, and DF SCF calculations for $OgTs_6$ are collected in Table 1.

## 4 Results and Discussion

It can be seen that whereas our NR HF calculations predict an *unbound* $OgTs_6$ with an Ae of -5.54 eV, the relativistic Dirac-Fock calculations predict the molecule to be *bound* with the Ae of 9.49 eV. Thus the contribution of ~ 15 eV at the DF level of theory to the atomization energy of the superheavy $OgTs_6$ is quite *dramatic*. Moreover, the contribution of the Breit magnetic effects of ~ 14 eV to the Ae is equally *dramatic*. Therefore it is mandatory that relativistic effects for the Ae of the superheavy OgTs$_6$ (and similar systems) should be included via accurate Dirac-Fock calculations as the less rigorous calculations may not be able to account for the pronounced effects of relativity for systems of SHE. It can be concluded from the calculated energy for the titled reaction (ΔE) that the forward reaction is most favored at the relativistic DF and DFBG levels ( with ΔE of 6.26 and 6.33 eV) and is least favored at the NR level of theory with ΔE of 8.81 eV. So there are significant contributions of ~ -2.50 to the ΔE at each DF and DFBG level.

5. **Relativistic electronic structure, bonding, and molecular spinor energy levels of e$_{1g}$ and e$_{1u}$ symmetries of OgTs6**

There are 204 e$_{1g}$ and 206 e$_{1u}$ doubly occupied molecular spinors (MS) in the relativistic ground state closed-shell electron configuration for OgTs$_6$ with 820 electrons under the double symmetry group (D$_{2h}$*) which Dirac [25] uses for calculations for octahedral systems. We shall first discuss the 204 e$_{1g}$ MS's labeled as 1e$_{1g}$ to 204 e$_{1g}$ in the ascending order of energy and the lowest energy MS 1e$_{1g}$, with the energy of -8185.8495 au consists of 0.76 Og(1s) plus 0.24 of small component (SC) of Og(1s), which is the same as the atomic spinor Og(1s). The presence of SC is due to the relativistic effects only and its contribution to 2s…6d ASs of Og and Ts decreases from ~0.24 to 0.002. Similarly, the contribution of SC decreases from ~0.24 to 0.002 in the MS's with almost ~ 99 % contribution of the atomic spinors of the atoms Ts and Og. The next 3-fold degenerate Ms's 2 e$_{1g}$ −4 e$_{1g}$ with the energy of -7988.4464 au consist of 0.77 Ts(1s) plus 0.23 SC, which equals the atomic spinor (AS) Ts(1s). To continue, the degenerate MS's 172 e$_{1g}$ and 173 e$_{1g}$ with the energy of -1.8570 au, consist of 0.999

Og(6d-) plus 0.0001 SC. It implies that the MS's ( 2 $e_{1g}$ −173 $e_{1g}$ ) which consist of the atomic spinors of higher energy up to the Ts(6p+) and Og (6d-) are fully occupied. Therefore all the MS's ( 2 $e_{1g}$ −173 $e_{1g}$ ) of OgTs6 are core-like atomic spinors of the two atoms Ts and Og and are non-bonding. Furthermore, it turns out the MS's ( 174 $e_{1g}$ −204 $e_{1g}$ ) consist of higher energy AS's up to the Ts (7p +) and Og (7p+) (with almost zero SC contribution) and all these MS's are non- bonding. It appears that there are no bonding MS's in the $e_{1g}$ symmetry which has 204 Ms's ( occupied with 408 electrons). Any bonding in OgTs6 therefore would be due to the net bonding of the 206 $e_{1u}$ symmetry MS's of OgTs6.

Next, we discuss the remaining 206 $e_{1u}$ MS's (labeled 1$e_{1u}$ to 206$e_{1u}$) which are occupied by 412 electrons in the octahedral OgTs6. The lowest 3-fold degenerate MS's (1$e_{1u}$-3$e_{1u}$ ) with the energy of -7988.4464 au consist of 0.77 Ts(1s) plus 0.23 SC of Ts(1s). The MS's 4$e_{1u}$ to 33$e_{1u}$ consist of ~ 0.93-0.98 AS's (indicated in the parenthesis of the AS ) Ts(2p-), Og(2p-), Ts(2p+), Ts(3s), Og(3p-), Og(3p+), Ts(3p+), Og(3p+), plus ~0.23 SC as in an analogues manner discussed above for the case of $e_{1g}$ symmetry. It can be seen that the MS's (1$e_{1u}$ - 33 $e_{1u}$ ) are like the inner core AS's of the Ts and Og atoms. The inner core behavior of the AS's continues from the MS 34$e_{1u}$ to MS 197 $e_{1u}$ where each MS consists of ~0.99 to 0.998 of AS's ( indicated in parenthesis) Ts(3d-), Ts(3d+), Ts(4f-) (each with an energy of -1.1391 au), plus ~0.02 to 0.01 of SC. Next, we discuss the MS's arising from the valence ASs of the two atoms Ts and Og. It should be mentioned, that the MS 198$e_{1u}$ with the energy of -0.8307 au consists of 0.94 Og (7p-) and 0.03 Ts(7p.) and it is bonding due to the bond between AS's Ts (7p-) and Og(7p-). The next MS's 199 $e_{1u}$ and 200 $e_{1u}$ (each with the energy of -0.6441 au) are degenerate and consist of 0.95 Ts ($7p_+$) and 0.04 Og ($7p_-$) and are weakly bonding. The MS's $201e_{1u}$ with the energy of -0.6262 au consists of 0.99 Ts ($7p_-$) and 0.01 Og ($7p_-$) and is antibonding. The degenerate MS's 202 $e_{1u}$ and 203 $e_{1u}$ (each with the energy of -0.3898 au) consist of 0.65 Og ($7p_+$) and 0.32 Ts ($7p_+$) and both the MS's are very strongly bonding. The next MS 204 $e_{1u}$ with the energy of -0.2526 au consists of 0.99 Ts ($7p_+$) and is nonbonding. Finally, the MS's 205 $e_{1u}$ and 206 $e_{1u}$ (each with the energy of -0.2523 au) consist of 0.97 Ts ($7p_+$) and 0.03 Og ($7p_+$), and both these MS's are antibonding.

This completes the bonding analysis and spinor energy level structure of all the 410 MS's of both the $e_{1g}$ and $e_{1u}$ symmetries of the octahedral $OgTs_6$ which consists of seven superheavy elements with 820 electrons. To sum up, almost all the MS's are occupied by the core-like AS's of the two superheavy atoms Ts and Og, except a few bonding and antibonding MS's which arise by an interaction of only the valence $7p_+$ and $7p_-$ AS's of the two superheavy Og and Ts atoms.

## 6. Nonrelativistic electronic structure, bonding, and molecular orbital energy levels of $OgTs_6$

Molecular orbital energy level structure and bonding obtained from our nonrelativistic HF calculations are quite similar to those discussed in section 5 above for our DF relativistic calculations. We shall also use the double group symmetry notation even though the single group is appropriate for NR HF calculations. The number of MO's in the $e_{1g}$ and $e_{1u}$ symmetries is the same as in the case of relativistic DF treatment, viz, 204 $e_{1g}$ and 206 $e_{1u}$, respectively. Within each symmetry the MO's are labeled as $1e_{1g}$ … 204 $e_{1g}$, and as $1e_{1u}$…$206e_{1u}$ in the ascending order of energy. It turns out that in $e_{1g}$ symmetry, the $1e_{1g}$ MO with the lowest energy of -6221.1856 au consists of 1.00 NR AO Og(1s). The MO's $2e_{1g}$–$4e_{1g}$ are degenerate with the energy of -6112.3802 au and arise from the combination of the Ts(1s) AO's of the Ts ligands. However, the MO's $5e_{1g}$–$191e_{1g}$ (with orbital energy of -1144.6864 to -1.7996 au) consist of the core-like the AO's Ts(2s) ..Ts(6d) and Og(2s) ….Og(6d). All the MO's arising from the core-like AO's are nonbonding. The next MO $192e_{1g}$ with energy of -0.9374 au consists of 0.86 Og(7s), 0.08 Ts(7s), and 0.05 Ts(7p) is the strongest bonding MO of the $e_{1g}$ symmetry. The doubly degenerate MO's $193e_{1g}$ –$194e_{1g}$ (each with the energy of -0.6926 au) consist of 0.98 Ts(7s) and 0.01Og(6d) and are non-bonding core-like AO's of Ts(7s). The next MO $195e_{1g}$ (with the energy of -0.6750 au) consists of 0.92 Ts(7s) and 0.08 Og(7s) is non-bonding as well. The next three degenerate MO's, $196e_{1g}$-$198e_{1g}$ (each with an energy of -0.3273 au) consist of 0.96 Ts(7p) and 0.03 Og(6d) are weakly bonding. The MO's $199e_{1g}$-$203\ e_{1g}$ ( each with the energy of -0.3273 au) are degenerate and each MO consists of 1.00 Ts(7p) AO. Finally, the MO 204 $e_{1g}$ (with the orbital energy of -0.2793 au) consists of 0.95 Ts(7p) and 0.05 Og (7s) and is anti-bonding.

There are 206 MO's of $e_{1u}$ symmetry and the three lowest energy degenerate MO's $1e_{1u}$-$3e_{1u,}$ ( each with the energy of -6112.3801 au ) consist of core-like pure Ts(1s) AO's and are nonbonding MO's. The next three degenerate MO's $4e_{1u}$-$6e_{1u}$ (each with an energy of -1122.5658 au) consist of core-like Ts (2s) AO's and are nonbonding. The nonbonding MO's $7e_{1u}$-$194e_{1u}$ consist of core-like pure AO's Og(2p) to Og(6p) and Ts(2p) to Ts(6d).We would like to mention that the degeneracy of the MO's is quite high in this range of MOs and, as an example, we quote here the *9-fold* degenerate (10 $e_{1u}$ -18 $e_{1u}$) MO's with the energy of -1093.6925 au, which consist of core-like pure Ts(2p) AO's of the Ts ligands. An example of a higher degeneracy is the set of *15-fold* degenerate MO's ($122e_{1u}$ -$136e_{1u}$) with the energy of -15.6081 au, and all these MO's are nonbonding and consist of core-like pure Ts(5d) AO's. A *21-fold* degenerate set of non-bonding MO's ($144e_{1u}$ -$164e_{1u}$) with the energy of -7.2138 au arises from a combination of core-like pure Ts(5f) AO's of the Ts ligands. Another example of a *15-fold* degenerate set of non-bonding MO's ($180e_{1u}$ -$194e_{1u}$) with the energy of -1.7994 au arises from a combination of pure core-like Ts(6d) AO's. It turns out that all the 194 MOs ($1e_{1u}$ -$194e_{1u}$) are nonbonding and consist of core-like pure 1s-6p AO's of Og and 1s to 6d AO's of Ts. Therefore all the 194 MO's of OgTs$_6$ consist of core-like AO's of the Og and Ts atoms. The next 3-fold degenerate set of ($195e_{1u}$ -$197e_{1u}$) MO's with an energy of -0.7212 au consists of 0.83 Ts(7s), 0.13 Og(7p), 0.02 Ts(7p), 0.01 Ts(6d). This is the strongest bonding orbital and consists of valence orbitals of Og and Ts. The next set of 3-fold degenerate MO's ($198e_{1u}$ -$200e_{1u}$) with an energy of -0.5291au consists of 0.54 Og(7p), 0.26 Ts(7p), 0.17 Ts(7s), 0.02 Og(5f) and is bonding. Finally, there is a nonbonding 6-fold degenerate set of MO's (201 -206) with the energy of -0.3474 au, consisting of pure core-like AO with 0.992 Ts (7p). This completes the discussion of both the relativistic and nonrelativistic electronic structure, bonding, and molecular energy levels of OgTs$_6$.

## 7. Conclusion

We have performed the first *ab initio* all-electron fully relativistic Dirac–Fock and NR HF SCF calculations for the ground state of the superheavy organometallic octahedral Og(Ts)$_6$ and a summary of our major conclusions is as follows:

(1) Our relativistic DF SCF calculation predicts the superheavy octahedral Og(Ts)$_6$ to be *bound*, with the calculated DF atomization energy of 9.93 eV at the optimized Og-Ts bond distance of 3.35 Å; however, our NR HF SCF calculation at the optimized bond distance Og-Ts of 3.44 Å predicts the superheavy octahedral Og (Ts)$_6$ to be *unbound* with the calculated NR atomization energy of -5.54 eV at the optimized bond distance Og-Ts of 3.44 Å.

(2) Relativistic effects of ~15.6 eV to the atomization energy are *dramatic* and must be calculated using *ab initio* DF SCF or better methodology

(3) Our *ab initio* all-electron fully relativistic Dirac–Fock (DF) and nonrelativistic Hartree-Fock (NR) calculations *predict* the DF and NR HF energy of the titled reaction Og +3Ts$_2$ → OgTs$_6$ as 5.89 and 8.81 eV, respectively. The relativistic effects of ~-3 eV to the energy of the titled reaction are very significant and proper and rigorous relativistic treatment is mandatory for calculating the relativistic effects in chemical reactions involving superheavy systems

(4) There are very large relativistic corrections to the binding energies of the MOs, especially, the inner core orbitals of Og (Ts)$_6$. Moreover, very large S–O splitting is calculated for the core MOs that consist of the inner (core) p, d, and f AOs of Og as well as Ts atoms as expected.

(5) The 1s...7s AS's of the Og atom as well as the 1s… 7s AS's of the six Ts ligands, and their associated electrons are not involved in bonding in Og(Ts)$_6$, since they remain as if in pure AS's of Og and Ts atoms. Therefore, these core electrons could be treated in molecular calculations on compounds of the superheavy elements (SHE) Og and Ts, using appropriate frozen core or pseudopotential approximations with tremendous savings in computational cost.

(6) The predicted DF and NR Og-Ts bond distances are 3.35 and 3.44 Å, respectively and therefore the relativistic effects are not significant for calculation of the bond distances of superheavy systems like $OgTs_6$.

In conclusion, *ab initio fully relativistic all-electron* Dirac-Fock SCF calculations for molecular systems of SHE with about 800 electrons are no longer the *bottlenecks* of relativistic quantum chemistry of SHE.

**Acknowledgments**: This research used in part resources of the National Energy Research Scientific Computing Center (NERSC), which is supported by the Office of Science of the U.S.Department of Energy under Contract No.DE-AC03-76SF00098. We gratefully acknowledge the superb NERSC facility which is a *sine qua non* for our gargantuan calculations. Part of our extensive calculations was carried out using the Westgrid computing resources at Simon Fraser University, Burnaby, BC, Canada which are gratefully acknowledged.

**Table I Calculated** total energy (E in au), atomization energy (Ae in eV), bond distance (R in Å) for $OgTs_6$ ($O_h$) and energy for the reaction ($\Delta E^X$ in eV) for $Og + 3Ts_2 \rightarrow OgTs_6$ predicted with our NR, DF, and DFBG calculations.

|  | $OgTs_6$ |
|---|---|
| $E^{NR}$ | -318700.4110 |
| $E^{DF}$ | -375772.6361 |
| $E^{DFBG}$ | -375028.2911 |
| $Ae^{NR}$ | -5.54 |
| $Ae^{DF}$ | 9.49 |
| $Ae^{DFBG}$ | 9.37 |
| $R_{Og-Ts}^{DFBG}$ | 3.36 |
| $R_{Og-Ts}^{DF}$ | 3.35 |
| $R_{Og-Ts}^{NR}$ | 3.44 |
| $\Delta E^{NR}$ | 8.81 |
| $\Delta E^{DF}$ | 6.33 |
| $\Delta E^{DFBG}$ | 6.26 |